%% file: PAPER.tex
\pdfoutput=1
\documentclass[12pt,a4paper]{article}
\usepackage{ifthen} 
\newboolean{pdflatex}
\setboolean{pdflatex}{true} 

\newboolean{articletitles}
\setboolean{articletitles}{true} 

\newboolean{uprightparticles}
\setboolean{uprightparticles}{true} 

\newboolean{inbibliography}
\setboolean{inbibliography}{false} 

\input{preamble}
\usepackage{longtable} 

\begin{document}

\renewcommand{\thefootnote}{\fnsymbol{footnote}}
\setcounter{footnote}{1}

\input{title-LHCb-PAPER}

\renewcommand{\thefootnote}{\arabic{footnote}}
\setcounter{footnote}{0}

\pagestyle{plain}
\setcounter{page}{1}
\pagenumbering{arabic}

\section{Introduction} \label{intro}
One of the interesting features of \grpsuthree flavour symmetry breaking in the light neutral meson sector is the mixing of singlet and octet states in the pseudoscalar mesons. The physical \etaz and \etapr particles can be expressed as an admixture of light quark and strange quark  flavour eigenstates~\cite{DiDonato:2011kr},
\begin{equation}
  \left(\begin{array}{r} \ket{\etaz} \\ \ket{\etapr} \end{array} \right) = 
  \left(\begin{array}{rr} \cos\phi_{\rm p} & -\sin\phi_{\rm p} \\ \sin\phi_{\rm p} & \cos\phi_{\rm p} \end{array} \right)
  \left(\begin{array}{r} \ket{\etaz_{\quark}} \\ \ket{\etaz_{\squark}} \end{array} \right)\,,
  \label{mixing1} 
\end{equation}
where the quark states are defined as
\begin{equation} 
  \ket{\etaz_{\quark}} = \frac{1}{\sqrt{2}}\ket{\uubar + \ddbar}\quad\mbox{and}\quad\ket{\etaz_{\squark}}  = \ket{\ssbar}\,,
\end{equation}
and $\phi_{\rm p}$ is the mixing angle between light and strange quark states.
In principle all possible \grpsuthree flavour singlet states that contribute to the physical particles, including the gluonic wavefunction, \ket{gg}, and the heavier quarkonia wavefunctions, \ket{\ccbar} and \ket{\bbbar}, must be considered. Of these states, only \ket{gg} is considered to contribute to the mass eigenstates~\cite{Rosner:1982ey,Ahmady:1999tz,Shuryak:1997xd}, and this only occurs in the heavier \etapr meson~\cite{DiDonato:2011kr,Li:2007ky}. Introducing the gluon mixing angle $\phi_{\rm G}$, the \etapr wavefunction becomes 
\begin{equation} 
  \ket{\etapr} = \cos\phi_{\rm G}\sin\phi_{\rm p}\ket{\etaz_{\quark}} + \cos\phi_{\rm G}\cos\phi_{\rm p}\ket{\etaz_{\squark}} + \sin\phi_{\rm G}\ket{gg}\,.
\end{equation}
Many phenomenological and experimental studies have been carried out to determine the values of the mixing angles, and measurements are in the range $\phi_{\rm p}\approx38-46^{\circ}$~\cite{phiP1,phiP2,phiP3,phiP4,phiP5,phiP6,phiP7,phiP9,LHCB-PAPER-2012-022,phiPG1,phiPG2,phiPG3}. Refs.~\cite{phiPG1,phiPG2,phiPG3} also contain measurements of the gluonic mixing angle, which are consistent with zero, albeit with large uncertainties. Most recently, the mixing angles have been measured by the \lhcb experiment using \decay{\Bors}{\jpsi\etaorpr} decays, and are found to be $\phi_{\rm p}=(43.5^{+1.5}_{-2.8})^{\circ}$ and $\phi_{\rm G}=(0\pm25)^{\circ}$~\cite{LHCb-PAPER-2014-056}. 
\begin{figure}[t]
  \begin{center}
      \subfigure[\small Dominant electroweak loop diagram.]{
      \includegraphics[width=0.45\textwidth]{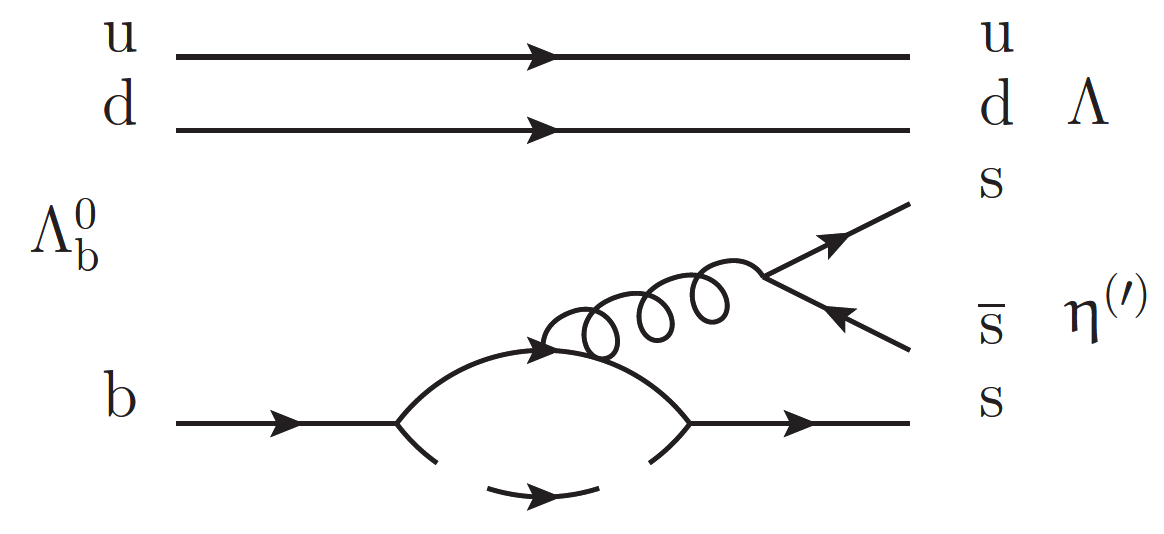}
      \label{B0_decay}
      }
    \subfigure[\small Non-spectator diagram.]{
      \label{B0_gluon_decay} 
      \includegraphics[width=0.42\textwidth]{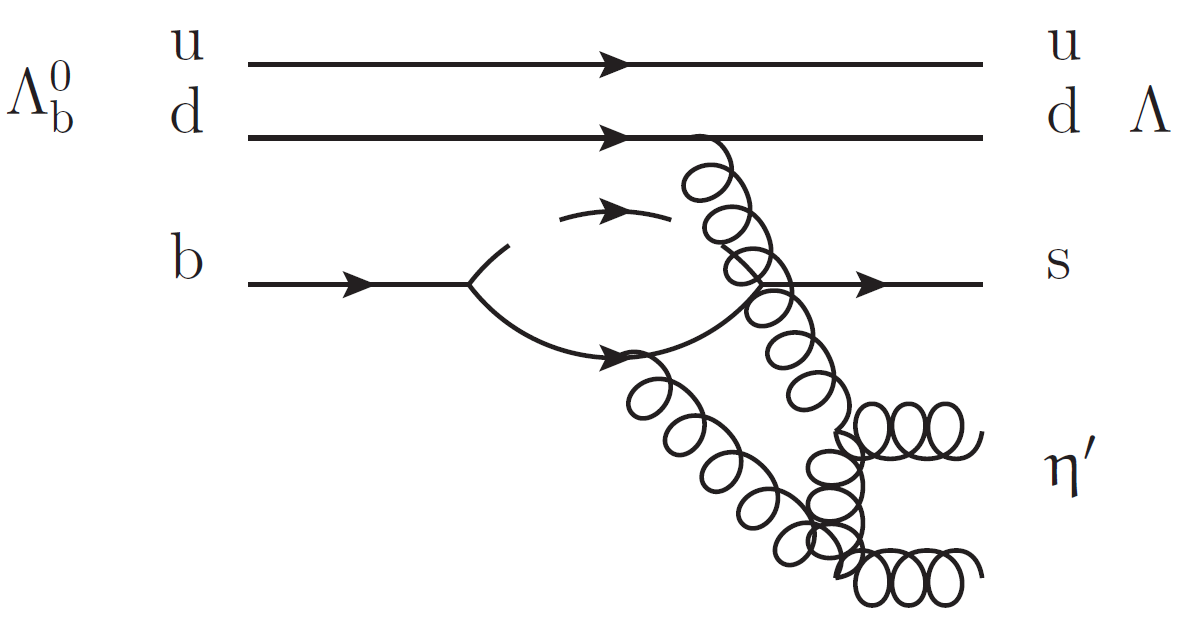}
    }
    \subfigure[\small Anomalous diagram.]{
        \includegraphics[width=0.45\textwidth]{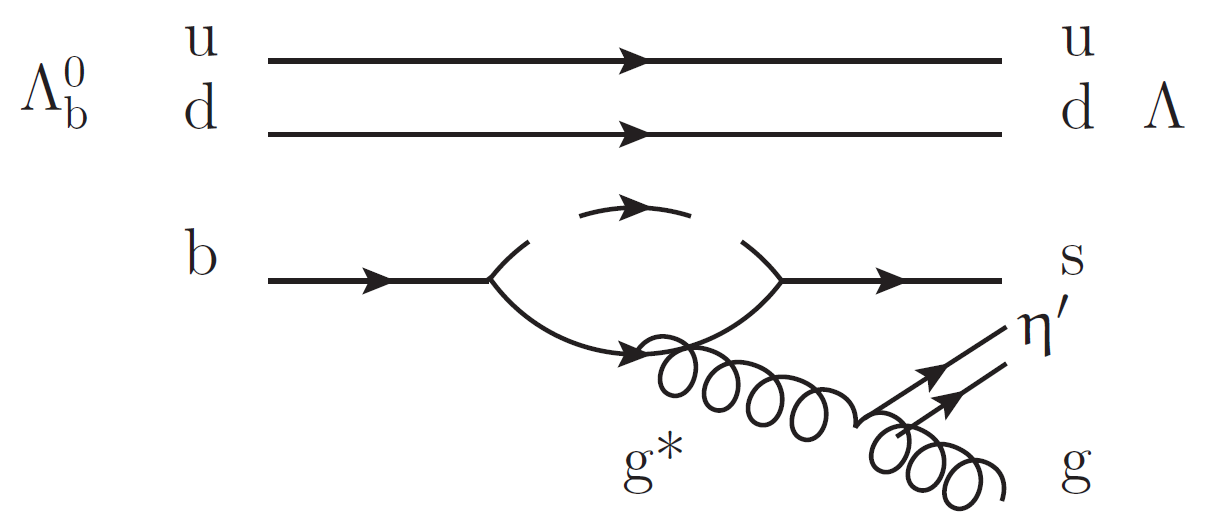}
        \label{B0_anomaly_decay}   
    }
    \caption{\small Feynman diagrams for the \LbEtaEtapL decay. The non-spectator and anomalous diagrams are available only to the \etapr through the gluonic contribution.}
    \label{B0_decays}
  \end{center}
\end{figure}

One consequence of the mixing is the difference in branching fractions for \bquark-hadron decays to final states containing \etaz and \etapr mesons. The gluonic mixing alters the branching fraction for the decays to \etapr mesons compared with the equivalent decay to \etaz mesons. Possible Feynman diagrams for amplitudes contributing to the \LbEtaEtapL decays are shown in Fig.~\ref{B0_decays}, with Fig.~\ref{B0_decay} showing the dominant \decay{\bquark}{\squark} transition via a penguin diagram. Due to the gluonic contribution of the wavefunction, extra Feynman diagrams of similar amplitude are available for the \etapr decay~\cite{Atwood:1997bn}. These include the possible non-spectator contribution, where the light quark radiates a gluon which can form an \etapr meson, as shown in Fig.~\ref{B0_gluon_decay}, and the anomalous contribution, where an excited gluon in the decay can radiate an \etapr meson, as shown in Fig.~\ref{B0_anomaly_decay}. The amplitudes for these extra processes can interfere to enhance or reduce the branching fraction to \etapr mesons. For example, the branching fraction for the \decay{\Bz}{\Kz\etapr} decay\footnote{Charge-conjugation is implied throughout.} has been measured to be $(6.6\pm0.4)\times10^{-5}$ which is over 50 times larger than the branching fraction for the \decay{\Bz}{\Kz\etaz} decay~\cite{PDG2014}. 
 
By measuring the relative branching fractions of many different decays to final states containing \etaz and \etapr mesons, it is possible to extract a measurement of the mixing angle $\phi_{\rm p}$. Decays of \bquark-baryons to final states containing \etaz or \etapr mesons have not yet been observed; however, the branching fractions of the \Lb decays have been estimated to be in the range $(1.8 - 19.0)\times10^{-6}$~\cite{Ahmady:2003jz}, depending on the model used to calculate the hadronic form factors. The interference between the anomalous and non-spectator contributions cancel in such a way that the branching fractions for the \LbEtapL and \LbEtaL decays are expected to be similar. Using the QCD sum rules approach to calculate the hadronic form factors, the branching fractions are predicted to be in the range $(6.0-19.0)\times10^{-6}$. The pole model approach predicts smaller branching fractions, in the range $(1.8-4.5)\times10^{-6}$.  For comparison, if the anomalous contribution is neglected, the branching fraction for the \LbEtapL decay would increase to $(33 - 40)\times10^{-6}$~\cite{Ahmady:2003jz}.

This paper describes the search for the \LbEtapL and \LbEtaL decays and measurement of the relative branching fractions with respect to the \decay{\Bz}{\Kz\etapr} decay, using the 3\invfb of data in $\proton\proton$ collisions collected in 2011 and 2012 by the \lhcb experiment.

\section{LHCb detector, trigger and event simulation}
The \lhcb detector~\cite{Alves:2008zz,LHCb-DP-2014-002} is a single-arm forward
spectrometer covering the \mbox{pseudorapidity} range $2<\eta <5$,
designed for the study of particles containing \bquark or \cquark
quarks. The detector includes a high-precision tracking system
consisting of a silicon-strip vertex detector surrounding the $\proton\proton$
interaction region~\cite{LHCb-DP-2014-001}, a large-area silicon-strip detector located
upstream of a dipole magnet with a bending power of about
$4{\rm\,Tm}$, and three stations of silicon-strip detectors and straw
drift tubes~\cite{LHCb-DP-2013-003} placed downstream of the magnet.
The tracking system provides a measurement of momentum, \ptot, of charged particles with
a relative uncertainty that varies from 0.5\% at low momentum to 1.0\% at 200\gevc.
The minimum distance of a track to a primary vertex, the impact parameter, is measured with a resolution of $(15+29/\pt)\mum$,
where \pt is the component of the momentum transverse to the beam, in \gevc.
Different types of charged hadrons are distinguished using information
from two ring-imaging Cherenkov (RICH) detectors~\cite{LHCb-DP-2012-003}. 
Photons, electrons and hadrons are identified by a calorimeter system consisting of
scintillating-pad and preshower detectors, an electromagnetic
calorimeter and a hadronic calorimeter. Muons are identified by a
system composed of alternating layers of iron and multiwire
proportional chambers~\cite{LHCb-DP-2012-002}.

The trigger~\cite{LHCb-DP-2012-004} consists of a
hardware stage, based on information from the calorimeter and muon
systems, followed by a software stage, in which all charged particles
with $\pt>500\ (300)\mevc$ are reconstructed for 2011 (2012) data. The software trigger requires a two-, three- or four-track
secondary vertex with a significant displacement from any primary
$\proton\proton$ interaction vertex~(PV). At least one charged particle
must have a transverse momentum $\pt > 1.7\gevc$ and be
inconsistent with originating from a PV.
A multivariate algorithm~\cite{BBDT} is used for
the identification of secondary vertices consistent with the decay
of a \bquark-hadron.
	
In the simulation, $\proton\proton$ collisions are generated using
\pythia~\cite{Sjostrand:2007gs} with a specific \lhcb
configuration~\cite{LHCb-PROC-2010-056}.  Decays of hadronic particles
are described by \evtgen~\cite{Lange:2001uf}, in which final-state
radiation is generated using \photos~\cite{Golonka:2005pn}. The
interaction of the generated particles with the detector, and its response,
are implemented using the \geant
toolkit~\cite{Allison:2006ve,*Agostinelli:2002hh} as described in
Ref.~\cite{LHCb-PROC-2011-006}.

\section{Event selection}
Candidate signal events are identified by reconstructing the \LbEtapL and \LbEtaL decays. The decay \BdEtapKs is used as a normalisation channel for the measurement of the branching fractions of the signal decays.

The long-lived \KS and \Lz particles are reconstructed through the \decay{\KS}{\pip\pim} and \decay{\Lz}{\proton}{\pim} decays. The reconstruction of these particles is labelled according to where in the LHCb detector the decay occurs. If the particle decay products produce hits in the vertex detector then the candidate is classified as \emph{Long (L)}; otherwise the candidate is referred to as \emph{Downstream (D)}. Since the track resolution is different for the two categories, the selection is optimised separately for \emph{L} and \emph{D} candidates.

The candidate \etapr mesons in the \LbEtapL decay are reconstructed as \EtapPiPiG and \EtapPiPiEta, where the \etaz meson decays into two photons. The candidate \etaz mesons used in the reconstruction of the \LbEtaL decay are reconstructed with the \EtaPiPiPi decay, with the \piz meson decaying into two photons. Since the charged tracks in the \EtaPiPiPi decay can be used for the trigger selection, the analysis is based on this mode rather than the neutral decays of the \etaz meson, which would have a lower efficiency.

Events that pass the trigger selection are subject to further requirements consisting of kinematic, particle identification and multivariate selections. A minimum requirement is placed on the fit quality of the reconstructed vertices. When reconstructing the \Bz and \Lb candidates, the direction angle\footnote{The direction angle is defined as the angle between the momentum vector of the particle and the vector between the PV and the decay vertex.} and the impact parameter are required to be consistent with the reconstructed particle originating from a PV, and the calculated lifetime should be significantly different from zero. The \Bz (\Lb) candidates are required to have $\pt>1.5\,(1.0)\gevc$.  The \KS mesons are required to have $\pt>1.2\gevc$ and a flight distance significantly different from zero. The invariant mass of the $\pip\pim$ pair should be within 14\mevcc of the known \Kz mass~\cite{PDG2014} for events in the \emph{L} category, and be within 23\mevcc for events in the \emph{D} category, where the mass window is chosen to be three times the resolution of the invariant mass. The \Lz baryon should have $\pt>1\gevc$ and an invariant mass within 15\mevcc of the known \Lz mass~\cite{PDG2014} for the \emph{L} category, and within 20\mevcc for the \emph{D} category. The \etapr and \etaz mesons from \Bz and \Lb decays are required to have $\pt>2\gevc$. Reconstructed $\pip\pim\gamma$ and $\pip\pim\etaz$ candidates are required to have an invariant mass in the range [0.9,1.05]\gevcc and $\pip\pim\piz$ candidates should have an invariant mass in a 150\mevcc window around the known \etaz meson mass~\cite{PDG2014}. Charged tracks are required to be of good quality with $\pt>300\mevc$. The \etaz mesons from \etapr decays, and \piz mesons from \etaz decays are reconstructed from two photons which can be resolved in the calorimeter, with an invariant mass in a 50\mevcc window around the \etaz or \piz mass~\cite{PDG2014}, and the \etaz or \piz mesons are required to have $\pt>200\mevc$.  Finally, photons are required to have a transverse energy $\et>200\mev$, and a large p-value for the single photon hypothesis, in order to reject background from misidentified \piz mesons where both decay photons form a single merged cluster in the calorimeter.

To improve the resolution of the reconstructed invariant mass, the full decay chain is refitted, where the tracks and displaced vertices are constrained, with the position of the PV of the \Bz or \Lb candidate fixed to the PV refitted using only tracks not associated to the \bquark-hadron decays, and the invariant mass of the \KS, \Lz, \etapr and \etaz particles fixed to their known masses~\cite{Hulsbergen:2005pu}. Candidates with a poor quality in this constrained fit are rejected, which removes approximately 90\% of background from the \emph{L} category and 20\% of the background from the \emph{D} category.

Information from the RICH detectors and the \lhcb calorimeter system is used in a neural network to construct a probability that a track is a pion, a kaon or a proton. This particle identification (PID) information is used to reduce the background from misidentified kaons, protons and pions.

A boosted decision tree (BDT) \cite{Breiman, Roe} is used to obtain further discrimination between signal and background events. A different BDT is trained for each signal channel and the normalisation channel, separately for the \emph{L} and \emph{D} categories and the 2011 and 2012 data samples. Samples of simulated events are used to model the signal decays, and events in data with a reconstructed invariant mass greater than 100\mevcc above the \Bz or \Lb mass~\cite{PDG2014} are used to model the combinatorial background. 

In this analysis, the numbers of candidates in both the signal and the background samples are limited and the performance of the BDT is improved in two ways. In the simulated signal sample no requirements are placed on the trigger selection. Instead, the kinematic distributions of the candidates are reweighted in order to match the harder distributions of candidates which pass the trigger selection. In addition, the signal and background samples are split in two, and two BDTs are trained for each channel. The first half of the sample is used to train a BDT which is applied to the second half of the data, and vice versa for the second BDT. This way, all data are available in the training of the BDTs, while any bias in the multivariate selection is avoided.

The BDT uses a set of variables with discriminating power between signal and background, including kinematic variables and vertex and track quality variables, and combines them into one variable which separates signal and background well. The selection is optimised using the figure of merit $\epsilon_{\mathrm{MVA}}$/$(\frac{a}{2}+\sqrt{N_{\rm B}})$~\cite{Punzi}, where $\epsilon_{\mathrm{MVA}}$ is the selection efficiency for a particular BDT selection, $a=3$ is the targeted signal significance in standard deviations, and $N_{\rm B}$ is the number of combinatorial background events reconstructed within the signal region and passing the BDT selection, found by performing an unbinned extended maximum likelihood fit to the sidebands of the data and extrapolating this fit into the signal region.

The final stage is to ensure that each event passing the selection contains exactly one \Bz or \Lb candidate. Due to the poor invariant mass resolution of the \etaorpr mesons, it is possible for low energy photons from the underlying event to be reconstructed with signal pions to create a second \etaorpr candidate. This happens in about $10\%-20\%$ of events which pass the selection. In this situation the candidate with the highest \et photon is kept, and all other candidates in the event are rejected.

The efficiency of this selection is $(0.032\pm0.001)\%$ in the \emph{L} category and $(0.030\pm0.001)\%$ in the \emph{D} category. This includes the efficiency of reconstructing and selecting events which are simulated within the \lhcb detector acceptance.

\section{Results}
\subsection{Fit results and signal yields}
An unbinned extended maximum likelihood fit to the candidate \bquark-hadron mass spectrum is performed on the data which pass the selection. The model used for the fit to both the signal and normalisation channels consists of an exponential function to describe the combinatorial background, and a sum of two Gaussian functions with a common mean to describe the signal. The ratio of the resolutions of the two Gaussian functions and the ratio of the signal yields are obtained from a fit to the mass distribution in simulated signal samples. These parameter values are used in the fit to the data, and only the resolution and signal yield of the first Gaussian function and, for the normalisation channel, the common mean of the two Gaussian functions are allowed to float. The resolution of the reconstructed $\KS\etapr$ invariant mass is $30.0\pm1.5\mevcc$ for \emph{L} candidates and $29.4\pm1.3\mevcc$ for \emph{D} candidates. For the signal channels, the resolution of the reconstructed $\Lz\etapr$ invariant mass is $29.1\pm1.8\mevcc$ ($31.1\pm4.5\mevcc$) for candidates reconstructed with \EtapPiPiG in the \emph{L}(\emph{D}) category, and $47.8\pm9.2\mevcc$ ($56.6\pm10.9\mevcc$) for candidates reconstructed with \EtapPiPiEta in the \emph{L} (\emph{D}) category. Reconstructed $\Lz\etaz$ candidates have an invariant mass resolution of $49.4\pm3.4\mevcc$ in the \emph{L} category and $47.6\pm9.2\mevcc$ in the \emph{D} category.  The parameters of the fit are found to be consistent between the 2011 and 2012 data samples, and so the two samples are added together to perform an overall fit.

The possible presence of physics backgrounds has been investigated. The most likely backgrounds are: \bquark-hadron decay modes to mesons with open charm and an \etaorpr meson, with a \piz meson which is not reconstructed; the nonresonant decays to \KS or \Lz particles with two charged pions which are combined with a combinatorial photon, \piz or \etaz meson to form an \etaorpr meson candidate; or, in the case of the \EtapPiPiG decays, the nonresonant \decay{\Bz}{\KS\pip\pim\g} or \decay{\Lb}{\Lz\pip\pim\g} decays. These backgrounds are rejected well by the BDT selection, and there are expected to be fewer than one candidate from each category passing the selection.

\begin{figure}[t]
  \begin{center}
      \includegraphics[width=0.48\textwidth]{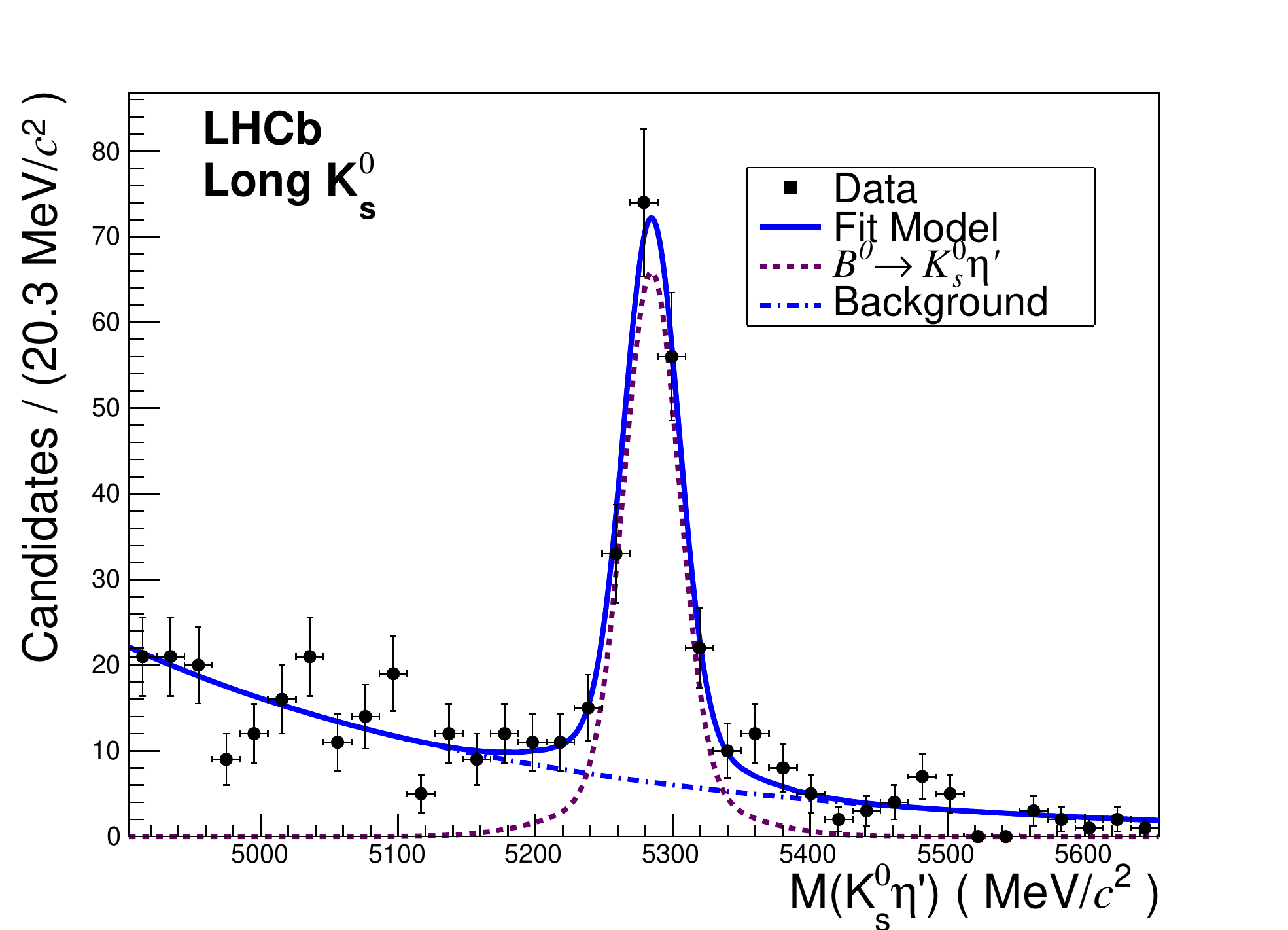}
      \includegraphics[width=0.48\textwidth]{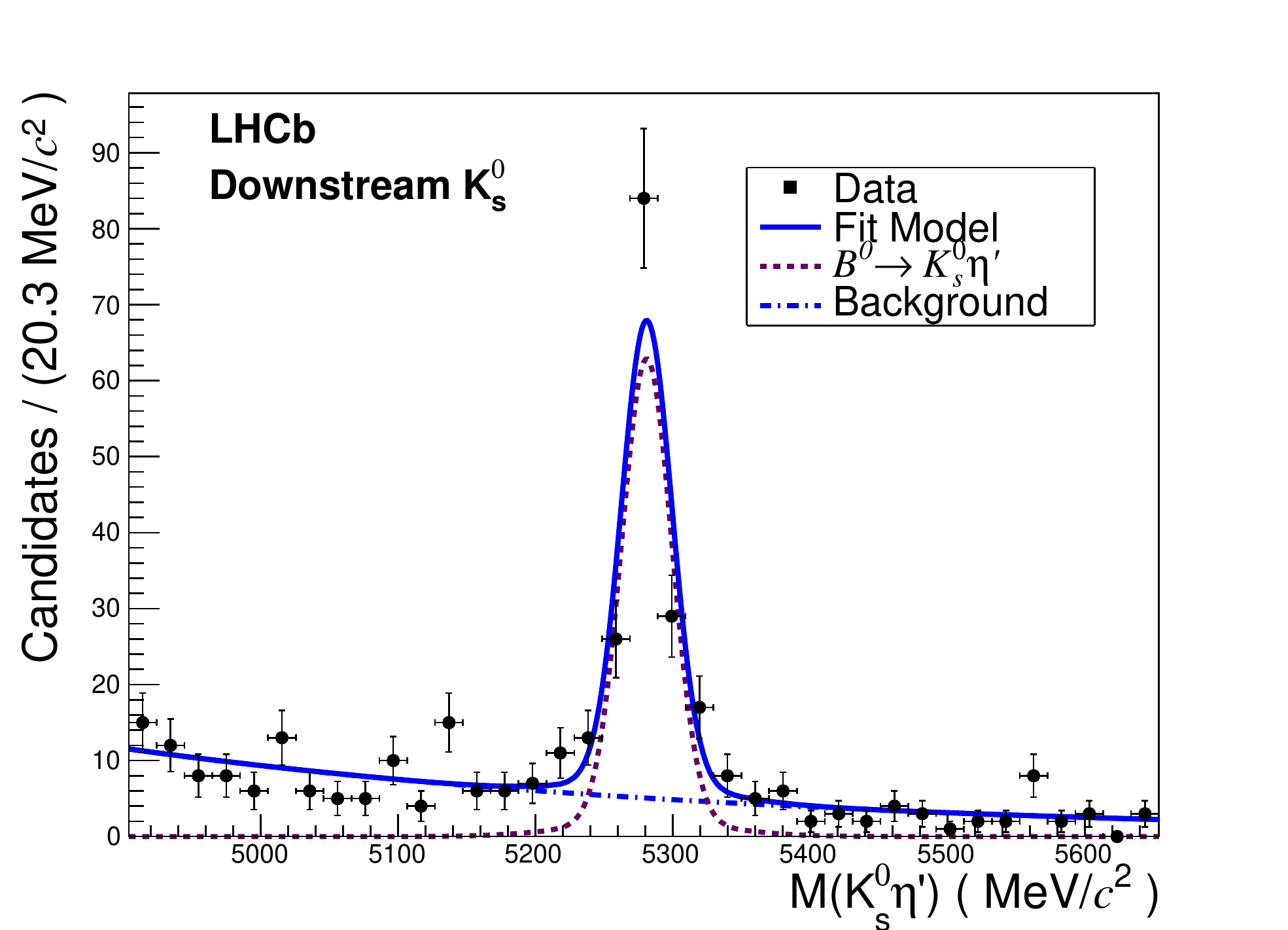}
  \end{center}
    \caption{\small Mass distribution of the selected \BdEtapKsPiPiG candidates in the 2011 and 2012 data, reconstructed in the \emph{L} (left) and \emph{D} (right) categories. The results of the fit, as described in the text, are overlaid.}
    \label{B0_mass_data_pvfit}
\end{figure}

\begin{figure}[t]
   \begin{center}
       \includegraphics[width=0.48\textwidth]{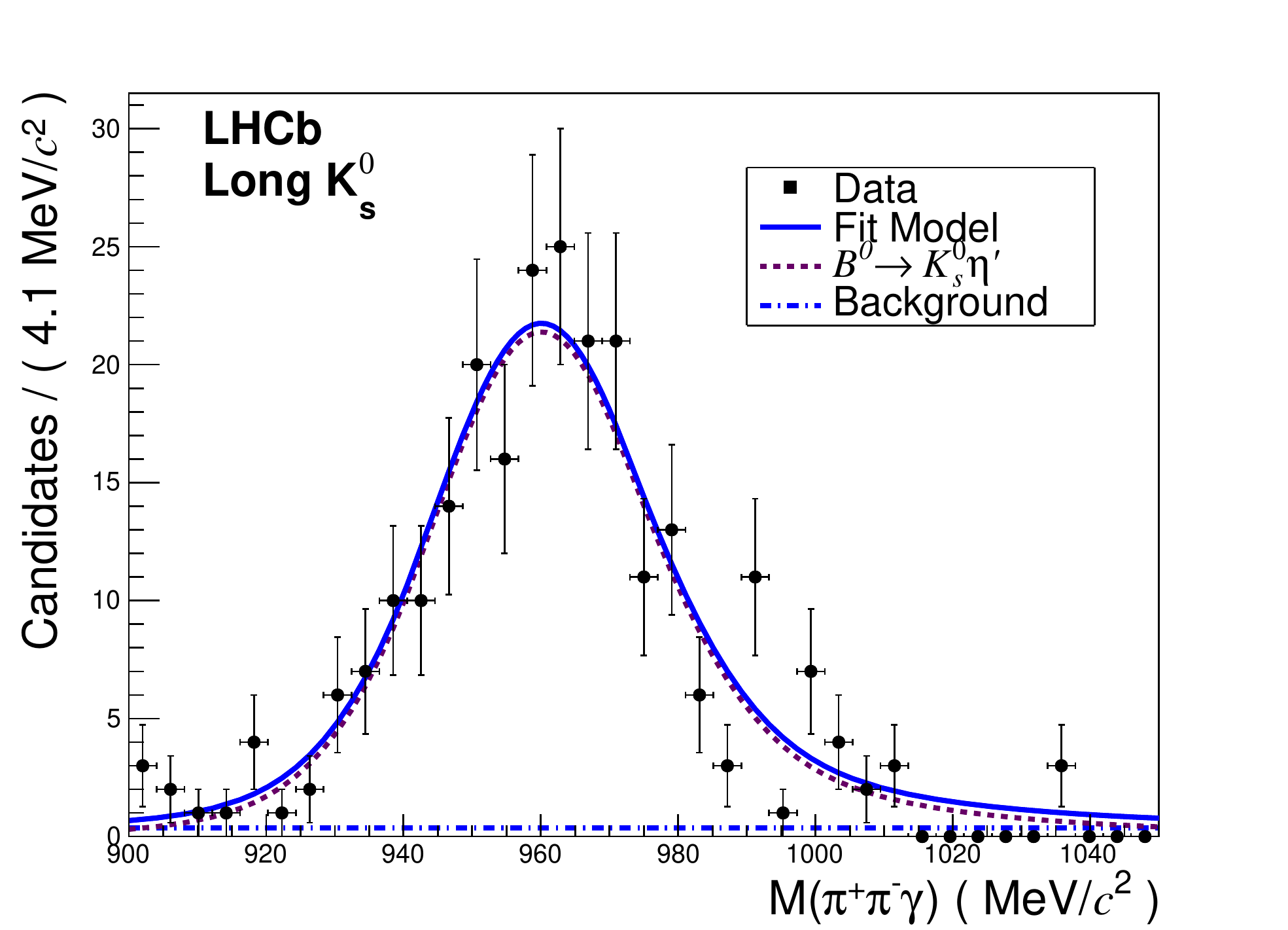}
       \includegraphics[width=0.48\textwidth]{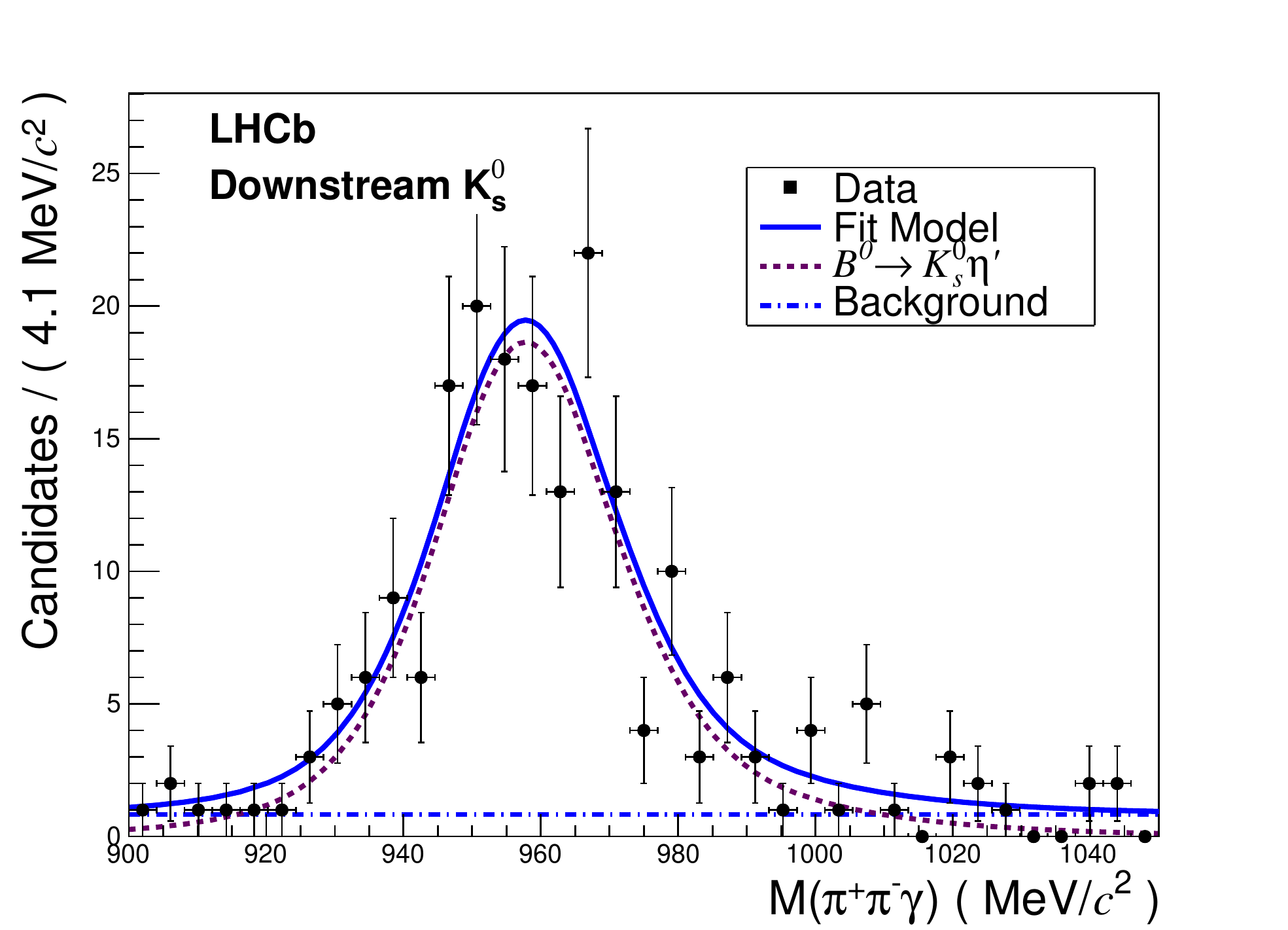}
   \end{center}
     \caption{\small Mass distribution of the reconstructed \etapr meson for selected \BdEtapKsPiPiG candidates, in the \emph{L} (left) and \emph{D} (right) categories. The results of the fit, as described in the text, are overlaid.}
     \label{etap_mass_data}
 \end{figure}

For the normalisation channel, the mass distribution of the selected $\KS\etapr$ candidates is shown in Fig.~\ref{B0_mass_data_pvfit}, with the result of the fit superimposed. The signal yields are $188\pm16$ \emph{L} candidates and $149\pm14$ \emph{D} candidates, and the signal to background ratio is 1.8 in the \emph{L} category, and 1.7 in the \emph{D} category. Fig.~\ref{etap_mass_data} shows the invariant mass distribution of the reconstructed \etapr mesons for these decays. The distribution is fitted with two Crystal Ball functions~\cite{Skwarnicki:1986xj}, and the parameters are found to be consistent with fits to the simulated samples, with a core resolution of $16\pm6\mevcc$ for \emph{L} candidates and $13\pm1\mevcc$ for \emph{D} candidates.

 \begin{figure}[t]
   \begin{center}
       \includegraphics[width=0.48\textwidth]{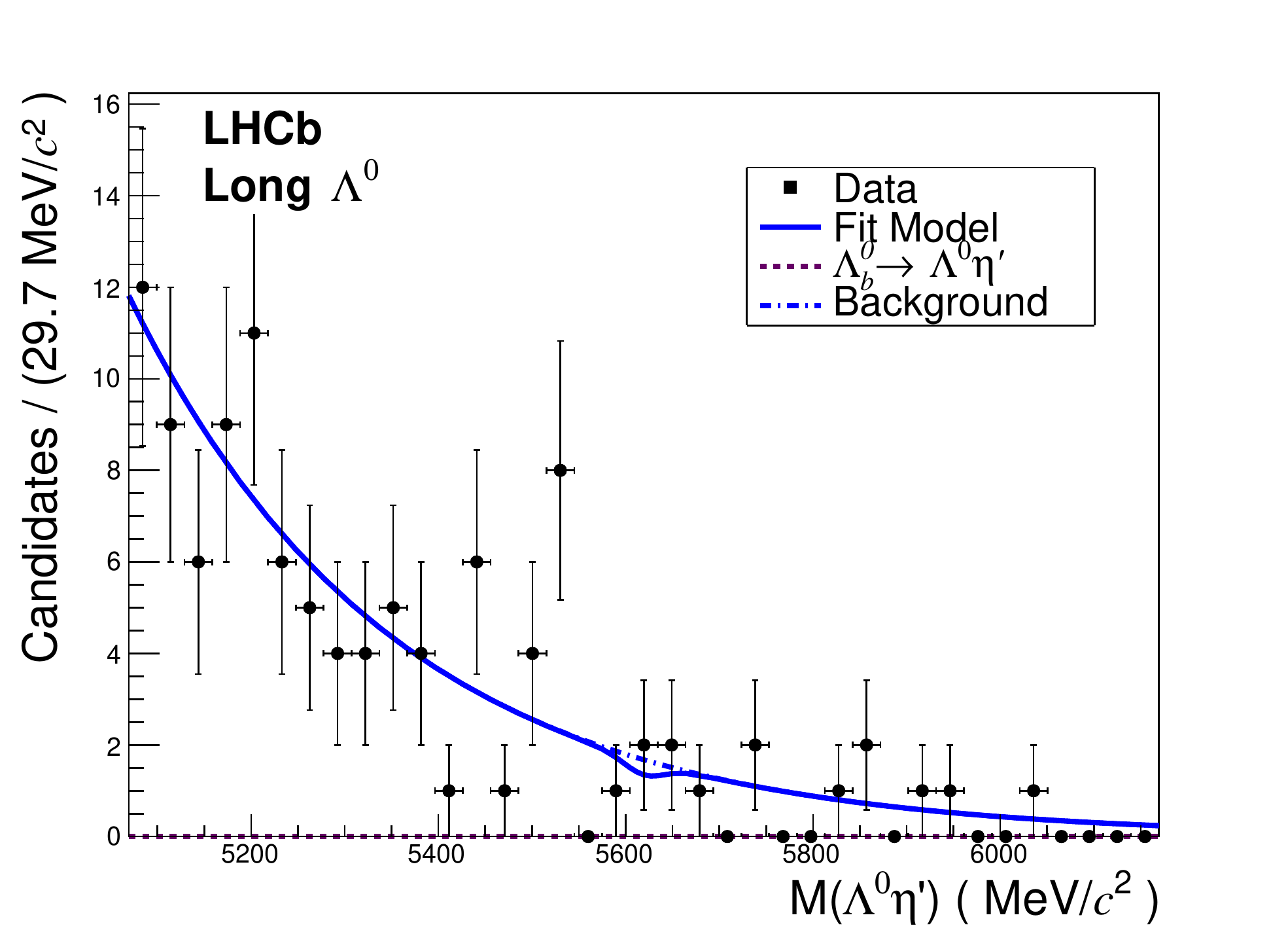}
       \includegraphics[width=0.48\textwidth]{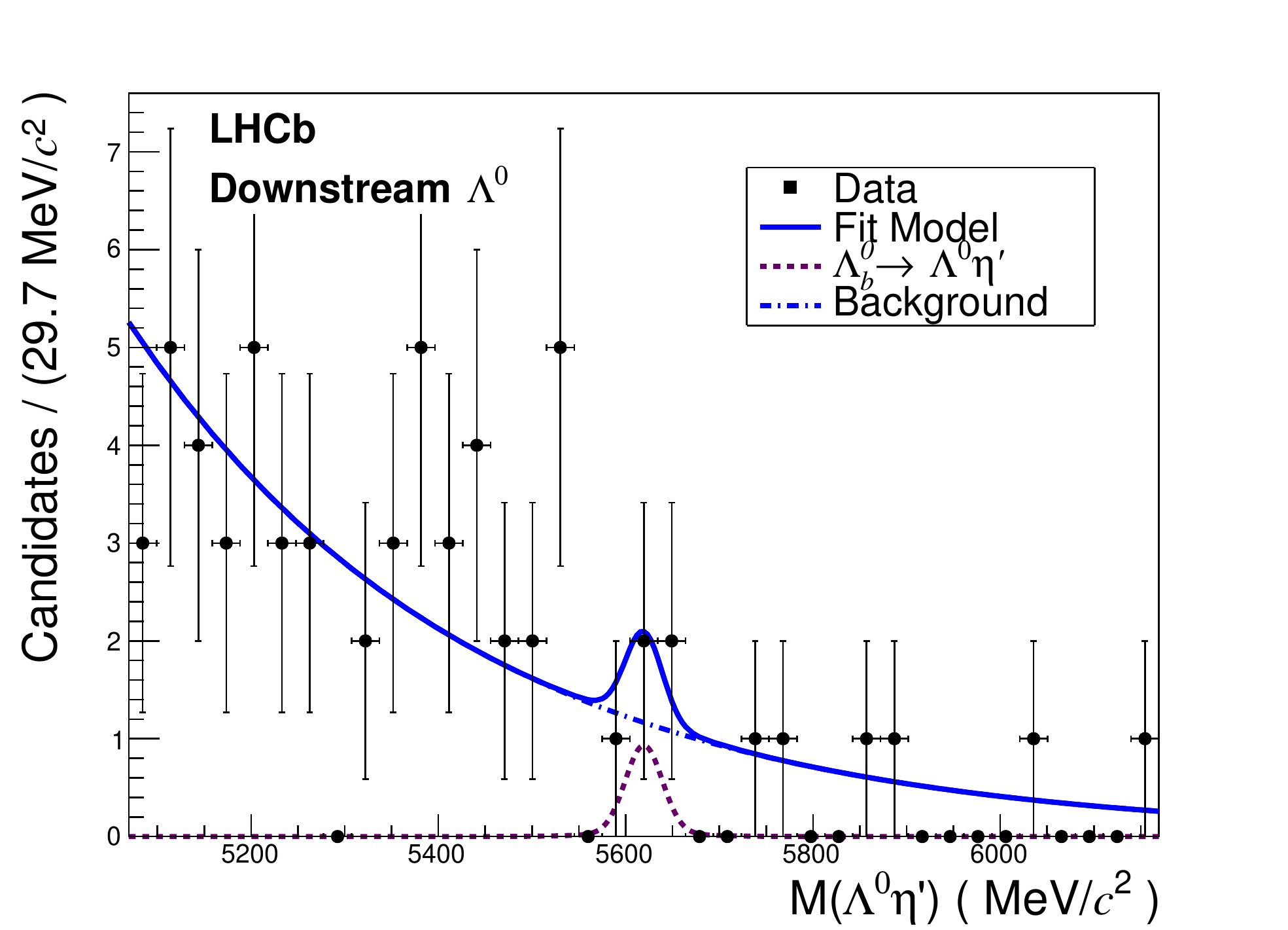}
   \end{center}
     \caption{\small Mass distribution of the selected \LbEtapLPiPiG candidates in the 2011 and 2012 data, reconstructed in the \emph{L} (left) and \emph{D} (right) categories. The results of the fit, as described in the text, are overlaid.}
     \label{Lb_mass_data_pvfit_1}
 \end{figure}
\begin{figure}[t]
   \begin{center}
       \includegraphics[width=0.48\textwidth]{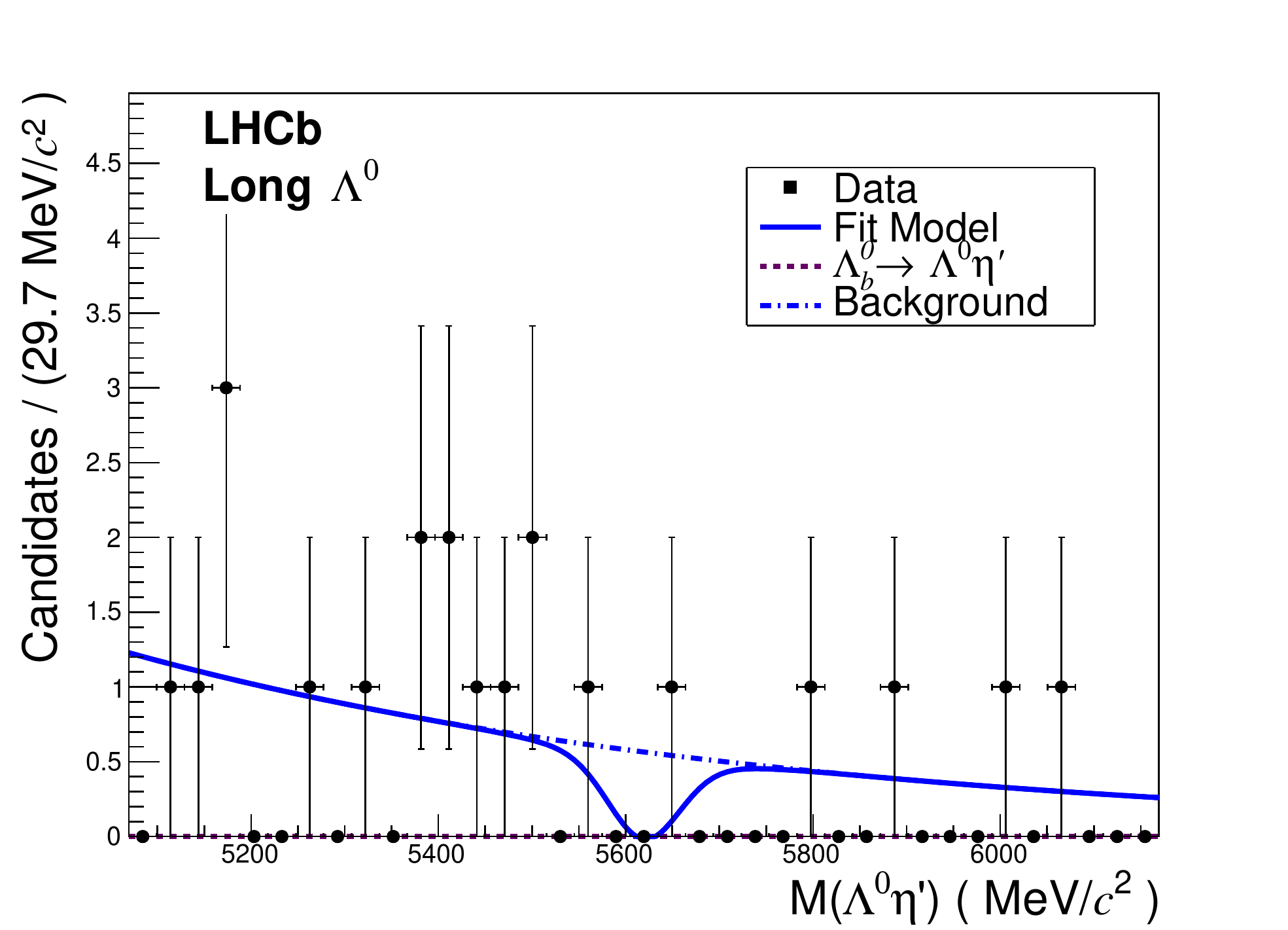}
       \includegraphics[width=0.48\textwidth]{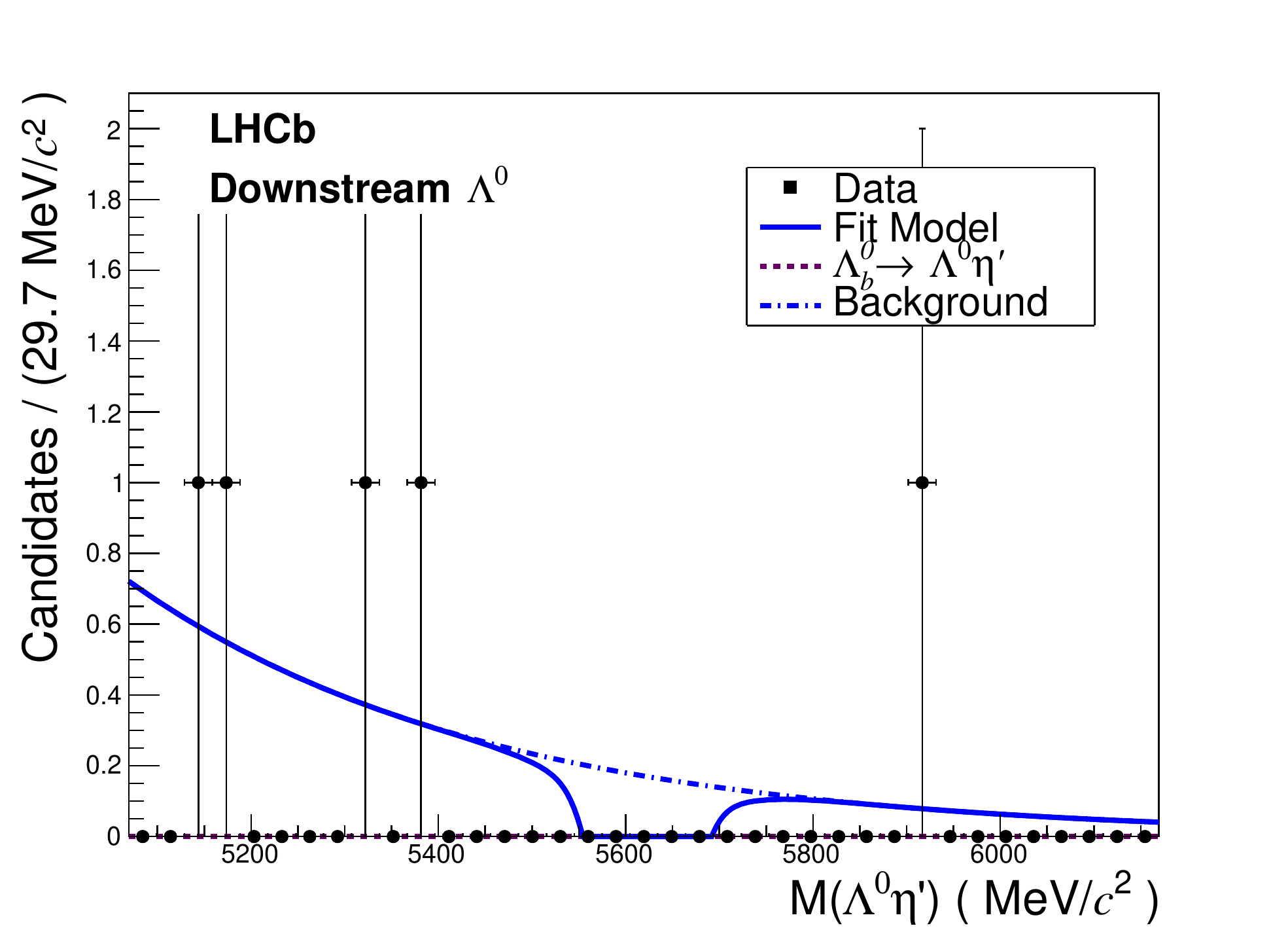}
   \end{center}
     \caption{\small Mass distribution of the selected \LbEtapLPiPiEta candidates in the 2011 and 2012 data, reconstructed in the \emph{L} (left) and \emph{D} (right) categories. The results of the fit, as described in the text, are overlaid.}
     \label{Lb_mass_data_pvfit_2}
 \end{figure}

 \begin{figure}[t]
   \begin{center}
       \includegraphics[width=0.48\textwidth]{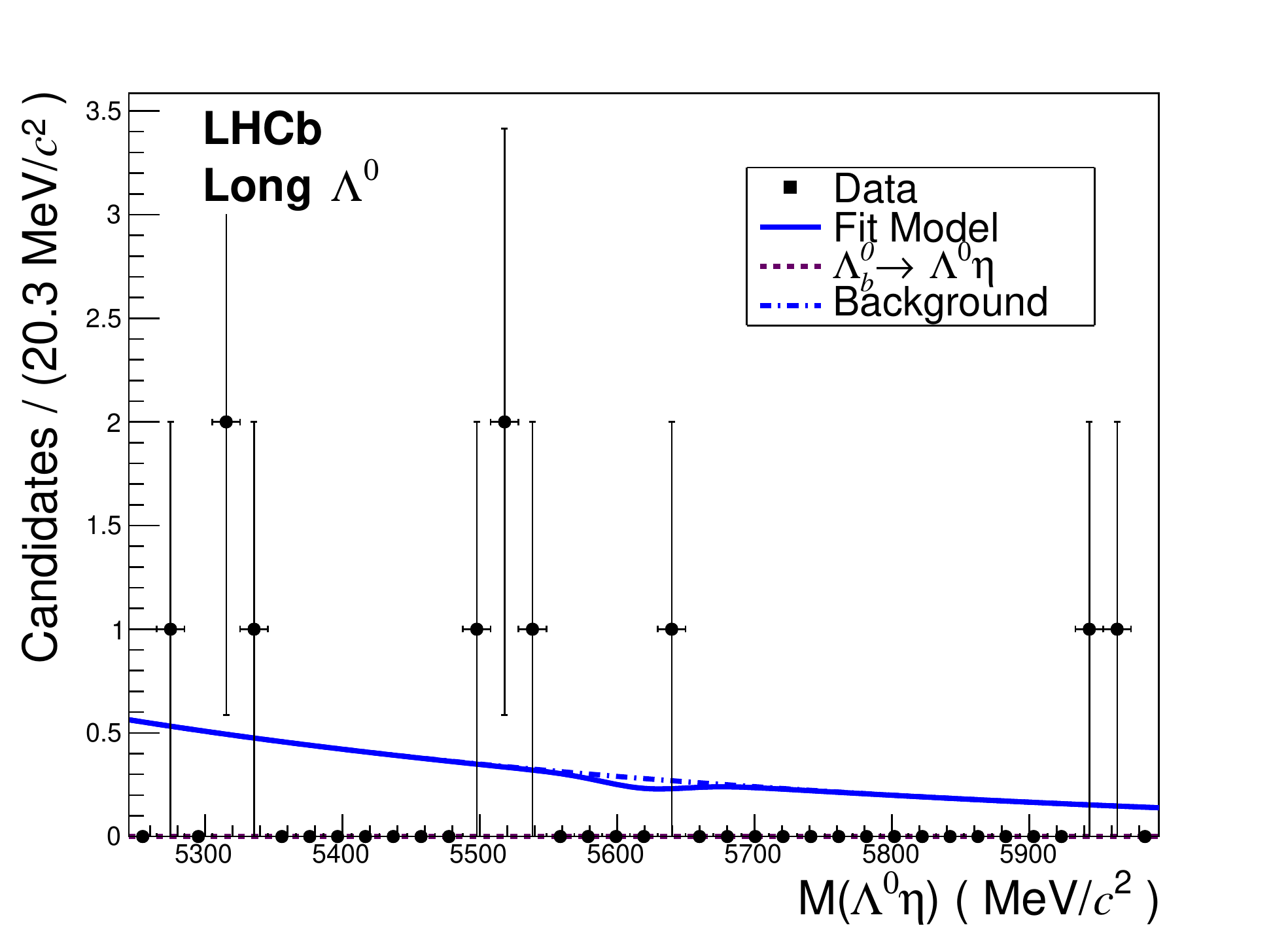}
       \includegraphics[width=0.48\textwidth]{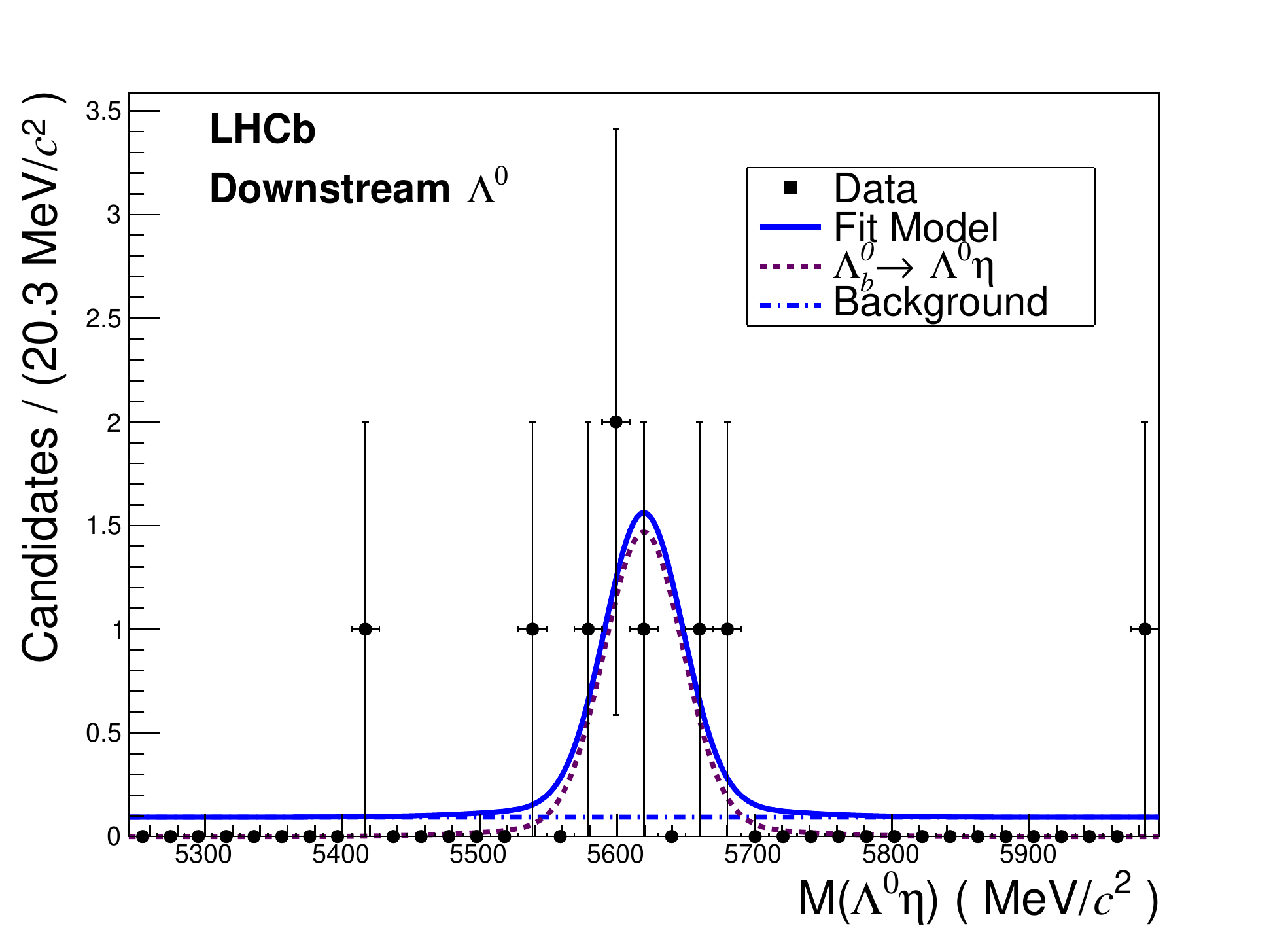}
   \end{center}
     \caption{\small Mass distribution of the selected \LbEtaLPiPiPiz candidates in the 2011 and 2012 data, reconstructed in the \emph{L} (left) and \emph{D} (right) categories. The results of the fit, as described in the text, are overlaid.}
     \label{Lb_mass_data_pvfit_3}
 \end{figure}

For the signal channel, Fig.~\ref{Lb_mass_data_pvfit_1} shows the $\Lz\etapr$ invariant mass distribution for candidates reconstructed in the \LbEtapLPiPiG decay, while Fig.~\ref{Lb_mass_data_pvfit_2} shows the same distribution for candidates in the \LbEtapLPiPiEta channel, and Fig.~\ref{Lb_mass_data_pvfit_3} shows the $\Lz\etaz$ invariant mass distribution for \LbEtaLPiPiPiz candidates. An unbinned extended maximum likelihood fit is performed using the same model as for the \Bz decay, with an exponential function to describe the combinatorial background and a sum of two Gaussian functions to model the signal; all parameters are fixed to the values found from fits to the simulation, and only the numbers of signal and background events are allowed to float in the fit. No significant signal is observed above the expected background for the \LbEtapL channel, and so an upper limit is placed on the ratio of branching fractions. From the fits, the combined signal yields are $1.0\pm4.4$ \LbEtapLPiPiG candidates and $-4.2\pm2.3$ \LbEtapLPiPiEta candidates. Evidence is seen for the presence of the \LbEtaL decay with a signal yield of $5.3\pm3.8$ candidates, and a significance of $3.0\sigma$.

\subsection{Branching fraction measurement}
The ratio of branching fractions can be measured for each signal decay with respect to the normalisation channel using
\begin{equation} 
\label{ratio}
\begin{split}
  R &\equiv \frac{\BF(\LbEtaEtapL)}{\BF(\BdEtapKz)} \\[0.3cm]
    &= \frac{N_{\rm S}(\Lb)}{N_{\rm S}(\Bz)} 
      \times \frac{\epsilon_{\mathrm{\rm tot}}(\Bz)}{\epsilon_{\mathrm{\rm tot}}(\Lb)}
      \times \frac{f_{\B}}{f_{\Lb}}
      \times \frac{1}{C_{\gamma}} 
      \times \frac{\BF(\etapr)}{\BF(\etaorpr)} 
      \times \frac{0.5\times\BF(\Kz)}{\BF(\Lz)} \\[0.3cm]
    &= \alpha \times \frac{N_{\rm S}(\Lb)}{N_{\rm S}(\Bz)} ,
\end{split}
\end{equation}

where: $N_{\rm S}$ is the number of signal events determined from the fits to data; $\epsilon_{\mathrm{\rm tot}}$ is the total efficiency, which is the product of the detector acceptance, reconstruction and selection efficiencies; $f_{\B}$/$f_{\Lb}$ is the ratio of \Bz to \Lb production fractions, previously measured by \lhcb~\cite{LHCb-PAPER-2014-004}; 1/$C_{\gamma}$ is a correction factor applied to account for the photon reconstruction efficiency in the \LbEtapLPiPiEta and \LbEtaLPiPiPiz decays, which have more photons in the decay than the normalisation channel; and $\BF(\etapr)/\BF(\etaorpr)$ and $\BF(\KS)/\BF(\Lz)$ are the ratios of \etapr to \etaorpr branching fractions and  \KS to \Lz branching fractions, respectively, from Ref.~\cite{PDG2014}, where the factor 0.5 accounts for the fact that only half the \Kz mesons decay as \KS mesons. Each of these factors have been measured, as described in Sec.~5, and are given in Table~\ref{alpha} along with the calculated value of $\alpha$ for each signal channel.

\begin{table}[tb]
  \begin{center}
  \label{alpha}
  \caption{\small Components contributing to the scale factor $\alpha.$}
      \begin{tabular}{c|cc|cc|cc} 
			&\multicolumn{2}{c|}{$\Lz\etapr(\pip\pim\g)$} & \multicolumn{2}{c|}{$\Lz\etapr(\pip\pim\etaz)$} & \multicolumn{2}{c}{$\Lz\etaz(\pip\pim\piz)$} \\			
    Factor   & \emph{L} & \emph{D} & \emph{L} & \emph{D} & \emph{L} & \emph{D} \\ \hline
    $\epsilon_{\rm tot}(\Bz)/\epsilon_{\rm tot}(\Lb)$ &  $2.3\pm0.1$ & $1.55\pm0.08$ & $7.4\pm0.6$ & $9.5\pm1.4$ & $4.6\pm0.3$ & $3.4\pm0.2$ \\    $f_{\B}/f_{\Lb}$                           &  \multicolumn{2}{c|}{$2.5\pm0.2$} &  \multicolumn{2}{c|}{$2.5\pm0.2$} &  \multicolumn{2}{c}{$2.5\pm0.2$} \\
    $1/C_{\gamma}$                         &  \multicolumn{2}{c|}{1 (fixed)} &  \multicolumn{2}{c|}{$0.95\pm0.04$} &  \multicolumn{2}{c}{$1.13\pm0.04$} \\
    $0.5\cdot\BF(\KS)/\BF(\Lz)$ &  \multicolumn{2}{c|}{$0.541\pm0.004$} &  \multicolumn{2}{c|}{$0.541\pm0.004$} &  \multicolumn{2}{c}{$0.541\pm0.004$}\\ 
    $\BF(\etapr)/\BF(\etaorpr)$ &  \multicolumn{2}{c|}{1 (fixed)} &  \multicolumn{2}{c|}{$1.71\pm0.05$} &  \multicolumn{2}{c}{$1.31\pm0.03$}\\ \hline 
    $\alpha$                                         & $3.1\pm0.3$ & $2.1\pm0.2$ & $17.7\pm2.3$ & $22.8\pm4.0$ & $9.5\pm1.2$ & $7.0\pm0.8$  \\  
    \end{tabular}
  \end{center}
\end{table}

\section{Systematic uncertainties}
\begin{table}[tb]
  \caption{\small Fractional systematic uncertainties, in percent, on the ratio of branching fractions.}
  \label{syst_err}
  \begin{center}
    \begin{tabular}{c|cc|cc|cc} 
			&\multicolumn{2}{c|}{$\Lz\etapr(\pip\pim\g)$} & \multicolumn{2}{c|}{$\Lz\etapr(\pip\pim\etaz)$} & \multicolumn{2}{c}{$\Lz\etaz(\pip\pim\piz)$} \\			
      Source                        & \emph{L} (\%) & \emph{D} (\%) & \emph{L} (\%) & \emph{D} (\%) & \emph{L} (\%) & \emph{D} (\%) \\ \hline
      \BF($V^{0}$)                       & \multicolumn{2}{c|}{0.73} & \multicolumn{2}{c|}{0.73} & \multicolumn{2}{c}{0.73}    \\ 
      \BF(\etaorpr)                       & \multicolumn{2}{c|}{0.0} & \multicolumn{2}{c|}{2.9} & \multicolumn{2}{c}{2.3}    \\ 
      $f_{\B}/f_{\Lb}$                       & \multicolumn{2}{c|}{8.0}   & \multicolumn{2}{c|}{8.0}  & \multicolumn{2}{c}{8.0}    \\ 
      Fit model                           & 1.7  & 0.3   & 1.7  & 0.3 & 1.7  & 0.3  \\
      Ratio of $\epsilon_{\mathrm{acc}}$      & \multicolumn{2}{c|}{2.0}   & \multicolumn{2}{c|}{1.8} & \multicolumn{2}{c}{1.9}  \\ 
      Ratio of $\epsilon_{\mathrm{sel}}$      & 5.9  & 5.3 & 8.6 & 14.5 & 7.5 & 5.9   \\ 
      $\epsilon_{\rm trig}$                       & \multicolumn{2}{c|}{1.0}   & \multicolumn{2}{c|}{1.0}  & \multicolumn{2}{c}{1.0}    \\ 
      $C_{\gamma}$                          & \multicolumn{2}{c|}{--}  & \multicolumn{2}{c|}{4.2} & \multicolumn{2}{c}{4.6} \\ 
      PID                                          & 2.1   & 0.9  & 2.0 & 1.0 & 2.0 & 1.0   \\ 
      Multiple Cand                  & 1.4   &  1.9  &  2.2  &  1.4  & 2.0  &  1.7 \\ \hline
      Total                             & 10.6  &  10.1  & 13.4 & 17.5 &  12.7 & 11.5   \\ 
    \end{tabular}
  \end{center}
\end{table}
A summary of systematic uncertainties is given in Table~\ref{syst_err}. The largest systematic uncertainty for this analysis is due to the limited knowledge of the ratio of production fractions, $f_{\B}/f_{\Lb}$, which was measured in Ref.~\cite{LHCb-PAPER-2014-004} as a function of the \Lb pseudorapidity. For the average \Lb pseudorapidity in our signal sample, the ratio $f_{\B}$/$f_{\Lb}$ has a value of $2.5\pm0.2$. 

The systematic uncertainty on the ratio of the branching fractions of the \KS, \Lz and \etaorpr decays is calculated from the average values from Ref.~\cite{PDG2014}. 

There is an uncertainty on the number of signal decays due to the fit model. To evaluate this uncertainty, the parameters that are fixed in the extended unbinned maximum likelihood fit are varied within the uncertainties obtained from the fit to the simulated samples, and the systematic uncertainty is the relative change in the yield obtained from the fit. This is found to be a small effect, 1.7\% for the \emph{L} model and 0.3\% for the \emph{D} model.

The selection efficiencies are calculated with independent simulated samples, produced with different trigger conditions. The uncertainty on the measured ratio of efficiencies is the statistical uncertainty due to the number of simulation events generated. In addition, there is a systematic uncertainty due to the measured efficiency of the trigger selection as the software trigger changed during the data taking in 2012. The early 2012 setup is not modelled in the simulated samples, and so it is assumed that the efficiency of this stage of the trigger selection is the same for both periods. The uncertainty due to this assumption is taken to be 5\%, which is consistent with differences in the efficiencies measured in analyses of similar decays. Since this data comprises $\sim20\%$ of the total data sample, the overall uncertainty on the trigger efficiency is 1\%.

The reconstruction efficiency of the photons cannot be determined using the simulated samples, and so a data-driven method is used to correct for this efficiency~\cite{LHCb-PROC-2015-009}. The correction factor, $C_{\gamma}$, is found in bins of photon \pt by comparing the relative yields of reconstructed \decay{\Bp}{\jpsi\Kstarp(\to\Kp\piz)} and \decay{\Bp}{\jpsi\Kp} decays. Since the \LbEtapLPiPiG decay contains the same number of photons as the normalisation channel, this correction cancels in the ratio of efficiencies. However, for the \LbEtapLPiPiEta and \LbEtaLPiPiPiz decays, there is an extra photon in the signal channel compared to the normalisation channel. The correction factor is therefore applied to these channels, and an uncertainty is introduced due to the limited size of the data sample used to measure it. 

The systematic uncertainty of the PID selection is estimated by the difference between the efficiency estimates obtained in data from calibration samples, and the efficiency extracted from the simulation after reweighting the PID variables to match the respective distribution observed in data.

Finally, there is an uncertainty due to the procedure for handling events with more than one candidate. The efficiency for selecting the best candidate using the procedure described earlier, and the fraction of events which contain more that one candidate are both estimated using simulated samples of signal decays. The estimated uncertainty on each of these quantities is around 10\%, which is chosen as a conservative uncertainty. This introduces an overall systematic uncertainty of 1.4\% to 2.2\% depending on the signal channel.

\section{Confidence Intervals}

The number of signal events observed in each of the signal decays can be used to place a limit on the ratio of branching fractions of the signal decay with respect to the normalisation channel. The unified approach method presented in Ref.~\cite{Feldman:1997qc} is used to place a limit on the branching fraction. This method constructs confidence intervals based on a likelihood ratio method using the probability of observing $N_{\rm obs}$ signal events for a given branching fraction. For this analysis the probability is assumed to follow a Gaussian distribution with a resolution of $\sigma=\sqrt{\sigma_{\rm syst}^{2} + \sigma_{\rm stat}^{2}}$, where $\sigma_{\rm syst}$ is the systematic uncertainty described above and $\sigma_{\rm stat}$ is the statistical uncertainty on the number of events observed.

The 68\% and 90\% confidence level (CL) intervals are obtained for the \LbEtaL and \LbEtapL decays respectively, by combining the likelihoods of each category. The weighted average of the observed ratio of branching fractions is calculated from the number of events observed in each signal channel with their uncorrelated systematic uncertainties and the values of $\alpha$ in Table~\ref{alpha}, and this is used to construct the confidence intervals. The combined \emph{L} and \emph{D} categories are used to construct confidence intervals for the \LbEtaL decay. The likelihoods from the \LbEtapLPiPiG and \LbEtapLPiPiEta decays are then combined, with the \emph{L} and \emph{D} combined, to give a limit on the branching fraction for the \LbEtapL decay. The limit on the ratio of branching fraction with respect to the normalisation channel is found to be 
\begin{equation*} \frac{\BF(\LbEtapL)}{\BF(\BdEtapKz)} < 0.047 \mathrm{\ at\ 90\%\ CL,} \end{equation*}
for the \LbEtapL decay, and for the \LbEtaL decay the 68\% CL intervals are
\begin{equation*} \frac{\BF(\LbEtaL)}{\BF(\BdEtapKz)} = 0.142^{+0.11}_{-0.08} . \end{equation*}
Multiplying by the known value of \BF(\BdEtapKz) gives a limit on the branching fraction of the \LbEtapL decay
\begin{equation*} \BF(\LbEtapL) < 3.1\times10^{-6} \mathrm{\ at\ 90\%\ CL,}  \end{equation*}
and the 68\% CL intervals for the \LbEtaL decay are
\begin{equation*} \BF(\LbEtaL) = (9.3^{+7.3}_{-5.3})\times10^{-6} . \end{equation*}

\section{Conclusions}

A search is performed for the \LbEtapL and \LbEtaL decays in the full dataset recorded by the \lhcb experiment during 2011 and 2012, corresponding to an integrated luminosity of 3\invfb. No significant signal is observed above background for the \LbEtapL decay, and some evidence is seen for the \LbEtaL at the level of $3\sigma$. The \BdEtapKz decay is used as a normalisation channel, so that a limit is placed on the ratio of \LbEtaEtapL branching fractions with respect to the \BdEtapKz branching fraction using the unified approach. With the known value of the \BdEtapKz branching fraction, the upper limit on the absolute branching fraction of the \LbEtapL decay is $\BF(\LbEtapL)<3.1\times10^{-6}$ at 90\% CL. The branching fraction of the \LbEtaL decay is $\BF(\LbEtaL)=(9.3^{+7.3}_{-5.3})\times10^{-6}$.

These values can be compared with the branching fractions calculated in Ref.~\cite{Ahmady:2003jz}, and given in Sec.~\ref{intro}. The predicted branching fractions depend strongly on the method used to calculate the hadronic form factors, and on the parameters used in the calculation, as discussed earlier. Our results favour the branching fractions calculated using the pole model to estimate the hadronic form factors. In addition, our results are inconsistent with the prediction for \mbox{\BF(\LbEtapL)} obtained by neglecting the anomalous contribution to the decay amplitude, indicating that a gluonic component of the \etapr wavefunction should be present.

\section*{Acknowledgements} 
\noindent We express our gratitude to our colleagues in the CERN
accelerator departments for the excellent performance of the LHC. We
thank the technical and administrative staff at the LHCb
institutes. We acknowledge support from CERN and from the national
agencies: CAPES, CNPq, FAPERJ and FINEP (Brazil); NSFC (China);
CNRS/IN2P3 (France); BMBF, DFG, HGF and MPG (Germany); INFN (Italy); 
FOM and NWO (The Netherlands); MNiSW and NCN (Poland); MEN/IFA (Romania); 
MinES and FANO (Russia); MinECo (Spain); SNSF and SER (Switzerland); 
NASU (Ukraine); STFC (United Kingdom); NSF (USA).
The Tier1 computing centres are supported by IN2P3 (France), KIT and BMBF 
(Germany), INFN (Italy), NWO and SURF (The Netherlands), PIC (Spain), GridPP 
(United Kingdom).
We are indebted to the communities behind the multiple open 
source software packages on which we depend. We are also thankful for the 
computing resources and the access to software R\&D tools provided by Yandex LLC (Russia).
Individual groups or members have received support from 
EPLANET, Marie Sk\l{}odowska-Curie Actions and ERC (European Union), 
Conseil g\'{e}n\'{e}ral de Haute-Savoie, Labex ENIGMASS and OCEVU, 
R\'{e}gion Auvergne (France), RFBR (Russia), XuntaGal and GENCAT (Spain), Royal Society and Royal
Commission for the Exhibition of 1851 (United Kingdom).

\addcontentsline{toc}{section}{References}
\setboolean{inbibliography}{true}
\bibliographystyle{LHCb}
\bibliography{LHCb-DP,main,LHCb-PAPER,LHCb-CONF,myBib}

\newpage
\input{LHCb_HD_authorlist_2015-03-24.tex}

\end{document}

%% file: preamble.tex
\usepackage{microtype}
\usepackage{lineno}
\usepackage{xspace} 
\usepackage{multirow}

\usepackage[pdftex]{graphicx}  
\usepackage{color}
\usepackage{colortbl}
\graphicspath{{./figs/}} 

\usepackage{amsmath} 
\usepackage{amssymb}
\usepackage{amsfonts}
\usepackage{upgreek}

\usepackage{subfigure}

\newcommand*\patchAmsMathEnvironmentForLineno[1]{%
\expandafter\let\csname old#1\expandafter\endcsname\csname #1\endcsname
\expandafter\let\csname oldend#1\expandafter\endcsname\csname
end#1\endcsname
 \renewenvironment{#1}%
   {\linenomath\csname old#1\endcsname}%
   {\csname oldend#1\endcsname\endlinenomath}%
}
\newcommand*\patchBothAmsMathEnvironmentsForLineno[1]{%
  \patchAmsMathEnvironmentForLineno{#1}%
  \patchAmsMathEnvironmentForLineno{#1*}%
}
\AtBeginDocument{%
\patchBothAmsMathEnvironmentsForLineno{equation}%
\patchBothAmsMathEnvironmentsForLineno{align}%
\patchBothAmsMathEnvironmentsForLineno{flalign}%
\patchBothAmsMathEnvironmentsForLineno{alignat}%
\patchBothAmsMathEnvironmentsForLineno{gather}%
\patchBothAmsMathEnvironmentsForLineno{multline}%
}

\usepackage{hyperref}    
\usepackage[all]{hypcap} 

\input{lhcb-symbols-def} 
\input{my-symbols-def} 

\usepackage{cite} 
\usepackage{mciteplus}

%% file: lhcb-symbols-def.tex
\def\lhcb {\mbox{LHCb}\xspace}

\def\MagUp {\mbox{\em Mag\kern -0.05em Up}\xspace}

\ifthenelse{\boolean{uprightparticles}}%
{
 
 \def\Pgamma      {\ensuremath{\upgamma}\xspace}

 \def\Peta        {\ensuremath{\upeta}\xspace}

 \def\Ppi         {\ensuremath{\uppi}\xspace}

 \def\Ppsi        {\ensuremath{\uppsi}\xspace}

 \def\PDelta      {\ensuremath{\Delta}\xspace}                 
 \def\PXi      {\ensuremath{\Xi}\xspace}                 
 \def\PLambda      {\ensuremath{\Lambda}\xspace}                 
 \def\PSigma      {\ensuremath{\Sigma}\xspace}                 
 \def\POmega      {\ensuremath{\Omega}\xspace}                 
 \def\PUpsilon      {\ensuremath{\Upsilon}\xspace}                 
 
 \def\PB      {\ensuremath{\mathrm{B}}\xspace}                 
                  
 \def\PD      {\ensuremath{\mathrm{D}}\xspace}

 \def\PJ      {\ensuremath{\mathrm{J}}\xspace}                 
 \def\PK      {\ensuremath{\mathrm{K}}\xspace}

 \def\Pb      {\ensuremath{\mathrm{b}}\xspace}                 
 \def\Pc      {\ensuremath{\mathrm{c}}\xspace}                 
 \def\Pd      {\ensuremath{\mathrm{d}}\xspace}

 \def\Pi      {\ensuremath{\mathrm{i}}\xspace}

 \def\Pp      {\ensuremath{\mathrm{p}}\xspace}                 
 \def\Pq      {\ensuremath{\mathrm{q}}\xspace}                 
                  
 \def\Ps      {\ensuremath{\mathrm{s}}\xspace}                 
                  
 \def\Pu      {\ensuremath{\mathrm{u}}\xspace}

}
{
 
 \def\Pgamma      {\ensuremath{\gamma}\xspace}

 \def\Peta        {\ensuremath{\eta}\xspace}

 \def\Ppi         {\ensuremath{\pi}\xspace}

 \def\Ppsi        {\ensuremath{\psi}\xspace}                 
                  
 \mathchardef\PDelta="7101
 \mathchardef\PXi="7104
 \mathchardef\PLambda="7103
 \mathchardef\PSigma="7106
 \mathchardef\POmega="710A
 \mathchardef\PUpsilon="7107
                  
 \def\PB      {\ensuremath{B}\xspace}                 
                  
 \def\PD      {\ensuremath{D}\xspace}

 \def\PJ      {\ensuremath{J}\xspace}                 
 \def\PK      {\ensuremath{K}\xspace}

 \def\Pb      {\ensuremath{b}\xspace}                 
 \def\Pc      {\ensuremath{c}\xspace}                 
 \def\Pd      {\ensuremath{d}\xspace}

 \def\Pi      {\ensuremath{i}\xspace}

 \def\Pp      {\ensuremath{p}\xspace}                 
 \def\Pq      {\ensuremath{q}\xspace}                 
                  
 \def\Ps      {\ensuremath{s}\xspace}                 
                  
 \def\Pu      {\ensuremath{u}\xspace}

}

\makeatletter
\ifcase \@ptsize \relax
  \newcommand{\miniscule}{\@setfontsize\miniscule{4}{5}}
\or
  \newcommand{\miniscule}{\@setfontsize\miniscule{5}{6}}
\or
  \newcommand{\miniscule}{\@setfontsize\miniscule{5}{6}}
\fi
\makeatother

\DeclareRobustCommand{\optbar}[1]{\shortstack{{\miniscule (\rule[.5ex]{1.25em}{.18mm})}
  \\ [-.7ex] $#1$}}

\def\g      {{\ensuremath{\Pgamma}}\xspace}

\def\quark     {{\ensuremath{\Pq}}\xspace}

\def\uquark    {{\ensuremath{\Pu}}\xspace}
\def\uquarkbar {{\ensuremath{\overline \uquark}}\xspace}
\def\uubar     {{\ensuremath{\uquark\uquarkbar}}\xspace}
\def\dquark    {{\ensuremath{\Pd}}\xspace}
\def\dquarkbar {{\ensuremath{\overline \dquark}}\xspace}
\def\ddbar     {{\ensuremath{\dquark\dquarkbar}}\xspace}
\def\squark    {{\ensuremath{\Ps}}\xspace}
\def\squarkbar {{\ensuremath{\overline \squark}}\xspace}
\def\ssbar     {{\ensuremath{\squark\squarkbar}}\xspace}
\def\cquark    {{\ensuremath{\Pc}}\xspace}
\def\cquarkbar {{\ensuremath{\overline \cquark}}\xspace}
\def\ccbar     {{\ensuremath{\cquark\cquarkbar}}\xspace}
\def\bquark    {{\ensuremath{\Pb}}\xspace}
\def\bquarkbar {{\ensuremath{\overline \bquark}}\xspace}
\def\bbbar     {{\ensuremath{\bquark\bquarkbar}}\xspace}

\def\pion   {{\ensuremath{\Ppi}}\xspace}
\def\piz    {{\ensuremath{\pion^0}}\xspace}

\def\pip    {{\ensuremath{\pion^+}}\xspace}
\def\pim    {{\ensuremath{\pion^-}}\xspace}

\def\kaon    {{\ensuremath{\PK}}\xspace}
  \def\Kbar    {{\kern 0.2em\overline{\kern -0.2em \PK}{}}\xspace}

\def\KorKbar    {\kern 0.18em\optbar{\kern -0.18em K}{}\xspace}
\def\Kz      {{\ensuremath{\kaon^0}}\xspace}

\def\Kp      {{\ensuremath{\kaon^+}}\xspace}

\def\KS      {{\ensuremath{\kaon^0_{\rm\scriptscriptstyle S}}}\xspace}

\def\Kstarp  {{\ensuremath{\kaon^{*+}}}\xspace}

\newcommand{\etaz}{\ensuremath{\Peta}\xspace}
\newcommand{\etapr}{\ensuremath{\Peta^{\prime}}\xspace}

  \def\Dbar    {{\kern 0.2em\overline{\kern -0.2em \PD}{}}\xspace}

\def\DorDbar    {\kern 0.18em\optbar{\kern -0.18em D}{}\xspace}

\def\B       {{\ensuremath{\PB}}\xspace}
\def\Bbar    {{\ensuremath{\kern 0.18em\overline{\kern -0.18em \PB}{}}}\xspace}

\def\BorBbar    {\kern 0.18em\optbar{\kern -0.18em B}{}\xspace}
\def\Bz      {{\ensuremath{\B^0}}\xspace}

\def\Bu      {{\ensuremath{\B^+}}\xspace}

\def\Bp      {{\ensuremath{\Bu}}\xspace}

\def\Bd      {{\ensuremath{\B^0}}\xspace}

\def\Bors      {{\ensuremath{\B^0_{(\squark)}}}\xspace}

\def\jpsi     {{\ensuremath{{\PJ\mskip -3mu/\mskip -2mu\Ppsi\mskip 2mu}}}\xspace}

  \def\Y#1S{\ensuremath{\PUpsilon{(#1S)}}\xspace}

\def\proton      {{\ensuremath{\Pp}}\xspace}

\def\Lz          {{\ensuremath{\PLambda}}\xspace}
\def\Lbar        {{\ensuremath{\kern 0.1em\overline{\kern -0.1em\PLambda}}}\xspace}
\def\LorLbar    {\kern 0.18em\optbar{\kern -0.18em \PLambda}{}\xspace}

\def\Lb      {{\ensuremath{\Lz^0_\bquark}}\xspace}

\def\BF         {{\ensuremath{\cal B}}\xspace}

\newcommand{\decay}[2]{\ensuremath{#1\!\to #2}\xspace}         

\def\to                 {\ensuremath{\rightarrow}\xspace}

\def\grpsuthree {{\ensuremath{\mathrm{SU}(3)}}\xspace}

\def\AT#1     {\ensuremath{A_{\mathrm{T}}^{#1}}\xspace}           

\def\C#1      {\ensuremath{\mathcal{C}_{#1}}\xspace}                       
\def\Cp#1     {\ensuremath{\mathcal{C}_{#1}^{'}}\xspace}                    
\def\Ceff#1   {\ensuremath{\mathcal{C}_{#1}^{\mathrm{(eff)}}}\xspace}        
\def\Cpeff#1  {\ensuremath{\mathcal{C}_{#1}^{'\mathrm{(eff)}}}\xspace}       
\def\Ope#1    {\ensuremath{\mathcal{O}_{#1}}\xspace}                       
\def\Opep#1   {\ensuremath{\mathcal{O}_{#1}^{'}}\xspace}                    

\newcommand{\ket}[1]{\ensuremath{|#1\rangle}}              

\newcommand{\tev}{\ifthenelse{\boolean{inbibliography}}{\ensuremath{~T\kern -0.05em eV}\xspace}{\ensuremath{\mathrm{\,Te\kern -0.1em V}}}\xspace}
\newcommand{\gev}{\ensuremath{\mathrm{\,Ge\kern -0.1em V}}\xspace}
\newcommand{\mev}{\ensuremath{\mathrm{\,Me\kern -0.1em V}}\xspace}
\newcommand{\kev}{\ensuremath{\mathrm{\,ke\kern -0.1em V}}\xspace}
\newcommand{\ev}{\ensuremath{\mathrm{\,e\kern -0.1em V}}\xspace}
\newcommand{\gevc}{\ensuremath{{\mathrm{\,Ge\kern -0.1em V\!/}c}}\xspace}
\newcommand{\mevc}{\ensuremath{{\mathrm{\,Me\kern -0.1em V\!/}c}}\xspace}
\newcommand{\gevcc}{\ensuremath{{\mathrm{\,Ge\kern -0.1em V\!/}c^2}}\xspace}
\newcommand{\gevgevcccc}{\ensuremath{{\mathrm{\,Ge\kern -0.1em V^2\!/}c^4}}\xspace}
\newcommand{\mevcc}{\ensuremath{{\mathrm{\,Me\kern -0.1em V\!/}c^2}}\xspace}

\def\mum  {\ensuremath{{\,\upmu\rm m}}\xspace}

\def\invfb   {\ensuremath{\mbox{\,fb}^{-1}}\xspace}

\def\gsim{{~\raise.15em\hbox{$>$}\kern-.85em
          \lower.35em\hbox{$\sim$}~}\xspace}
\def\lsim{{~\raise.15em\hbox{$<$}\kern-.85em
          \lower.35em\hbox{$\sim$}~}\xspace}

\def\ptot       {\mbox{$p$}\xspace}
\def\pt         {\mbox{$p_{\rm T}$}\xspace}
\def\et         {\mbox{$E_{\rm T}$}\xspace}

\def\evtgen     {\mbox{\textsc{EvtGen}}\xspace}

\def\geant      {\mbox{\textsc{Geant4}}\xspace}

\def\photos     {\mbox{\textsc{Photos}}\xspace}

\def\pythia     {\mbox{\textsc{Pythia}}\xspace}

\def\tell1  {TELL1\xspace}
\def\ukl1   {UKL1\xspace}



%% file: my-symbols-def.tex
\def\etaorpr {\ensuremath{\Peta^{(\prime)}}\xspace}

\def\BdEtapKs    {\mbox{\decay{\Bd}{\KS\etapr}}\xspace}

\def\BdEtapKz    {\mbox{\decay{\Bd}{\Kz\etapr}}\xspace}

\def\LbEtapL     {\mbox{\decay{\Lb}{\Lz\etapr}}\xspace}
\def\LbEtaL      {\mbox{\decay{\Lb}{\Lz\Peta}}\xspace}
\def\LbEtaEtapL  {\mbox{\decay{\Lb}{\Lz\etaorpr}}\xspace}

\def\EtaPiPiPi   {\mbox{\decay{\Peta}{\pip\pim\piz}}\xspace}
\def\EtapPiPiG   {\mbox{\decay{\etapr}{\pip\pim\gamma}}\xspace}
\def\EtapPiPiEta {\mbox{\decay{\etapr}{\pip\pim\Peta}}\xspace}

\def\BdEtapKsPiPiG {\BdEtapKs\mbox{(\EtapPiPiG)}\xspace}

\def\LbEtapLPiPiG {\LbEtapL\mbox{(\EtapPiPiG)}\xspace}
\def\LbEtapLPiPiEta {\LbEtapL\mbox{(\EtapPiPiEta)}\xspace}
\def\LbEtaLPiPiPiz {\LbEtaL\mbox{(\EtaPiPiPi)}\xspace}


%% file: title-LHCb-PAPER.tex
\begin{titlepage}
\pagenumbering{roman}

\vspace*{-1.5cm}
\centerline{\large EUROPEAN ORGANIZATION FOR NUCLEAR RESEARCH (CERN)}
\vspace*{1.5cm}
\noindent
\begin{tabular*}{\linewidth}{lc@{\extracolsep{\fill}}r@{\extracolsep{0pt}}}
\ifthenelse{\boolean{pdflatex}}
{\vspace*{-2.7cm}\mbox{\!\!\!\includegraphics[width=.14\textwidth]{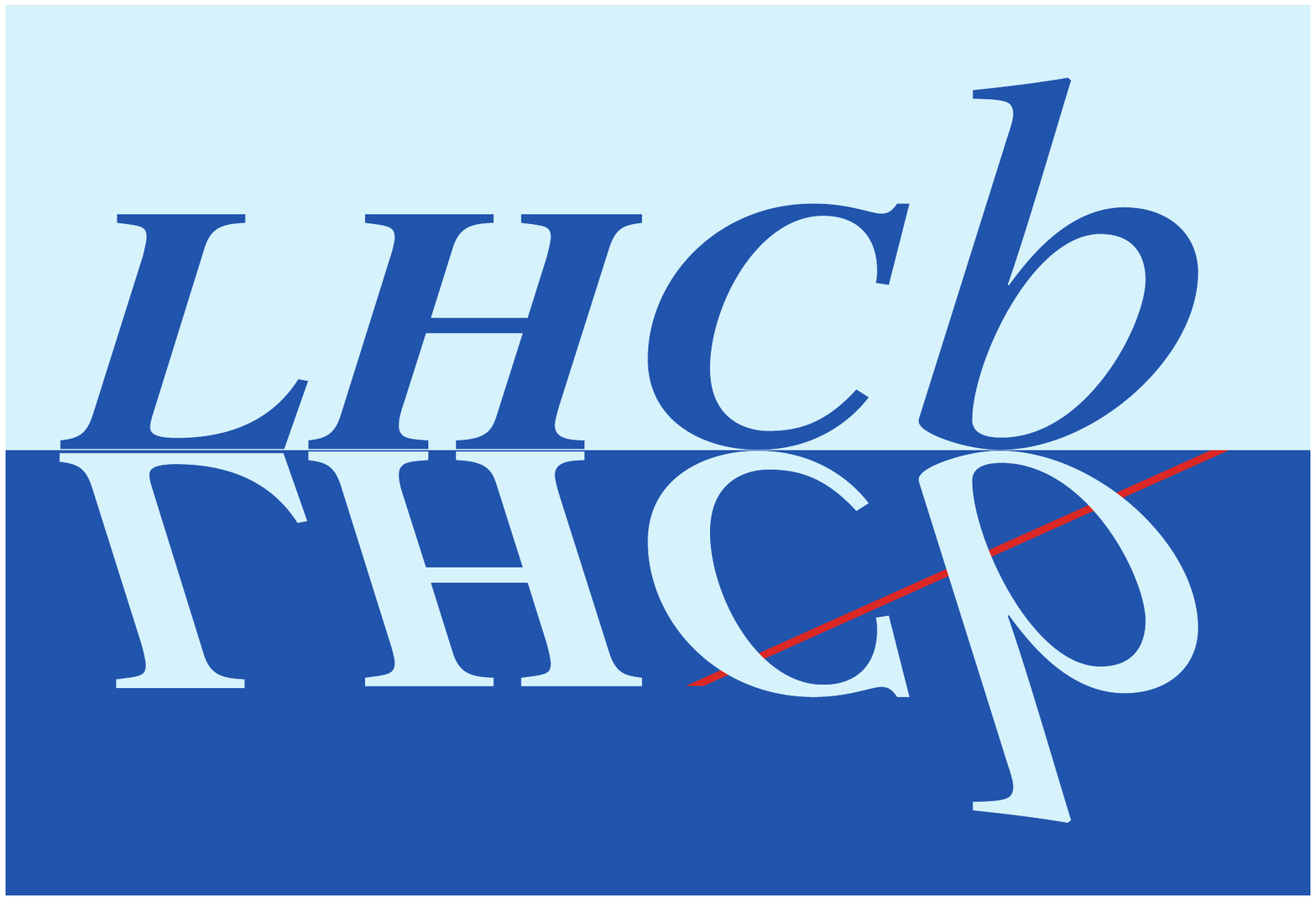}} & &}%
{\vspace*{-1.2cm}\mbox{\!\!\!\includegraphics[width=.12\textwidth]{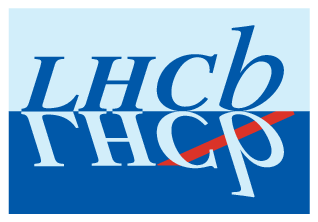}} & &}%
\\
 & & CERN-PH-EP-2015-112 \\  
 & & LHCb-PAPER-2015-019 \\   
 & & May 13, 2015 \\ 
 & & \\
\end{tabular*}

\vspace*{1.5cm}

{\bf\boldmath\huge
\begin{center}
	Search for the \LbEtapL and \LbEtaL decays with the \lhcb detector
\end{center}
}

\vspace*{0.5cm}

\begin{center}
The LHCb collaboration\footnote{Authors are listed at the end of this paper.}
\end{center}

\vspace{\fill}

\begin{abstract}
  \noindent
A search is performed for the as yet unobserved baryonic \LbEtapL and \LbEtaL decays with 3\invfb of proton-proton collision data recorded by the \lhcb experiment. The \BdEtapKs decay is used as a normalisation channel. No significant signal is observed for the \LbEtapL decay. An upper limit is found on the branching fraction of \mbox{$\BF(\LbEtapL)<3.1\times10^{-6}$} at 90\% confidence level. Evidence is seen for the presence of the \LbEtaL decay at the level of $3\sigma$ significance, with a branching fraction \mbox{$\BF(\LbEtaL)=(9.3^{+7.3}_{-5.3})\times10^{-6}$}.
\end{abstract}

\vspace*{1.5cm}

\begin{center}
 Published as JHEP 09 (2015) 006
\end{center}

\vspace{\fill}

{\footnotesize 
\centerline{\copyright~CERN on behalf of the \lhcb collaboration, licence \href{http://creativecommons.org/licenses/by/4.0/}{CC-BY-4.0}.}}
\vspace*{2mm}

\end{titlepage}

\newpage
\setcounter{page}{2}
\mbox{~}

\cleardoublepage

%% file: LHCb_HD_authorlist_2015-03-24.tex
\centerline{\large\bf LHCb collaboration}
\begin{flushleft}
\small
R.~Aaij$^{38}$, 
B.~Adeva$^{37}$, 
M.~Adinolfi$^{46}$, 
A.~Affolder$^{52}$, 
Z.~Ajaltouni$^{5}$, 
S.~Akar$^{6}$, 
J.~Albrecht$^{9}$, 
F.~Alessio$^{38}$, 
M.~Alexander$^{51}$, 
S.~Ali$^{41}$
% ...

{\footnotesize \it
% ...
$ ^{5}$Clermont Universit\'{e}, Universit\'{e} Blaise Pascal, CNRS/IN2P3, LPC, Clermont-Ferrand, France\\
$ ^{6}$CPPM, Aix-Marseille Universit\'{e}, CNRS/IN2P3, Marseille, France\\
% ...
$ ^{9}$Fakult\"{a}t Physik, Technische Universit\"{a}t Dortmund, Dortmund, Germany\\
% ...
$ ^{37}$Universidad de Santiago de Compostela, Santiago de Compostela, Spain\\
$ ^{38}$European Organization for Nuclear Research (CERN), Geneva, Switzerland\\
% ...
$ ^{41}$Nikhef National Institute for Subatomic Physics, Amsterdam, The Netherlands\\
% ...
$ ^{46}$H.H. Wills Physics Laboratory, University of Bristol, Bristol, United Kingdom\\
% ...
$ ^{51}$School of Physics and Astronomy, University of Glasgow, Glasgow, United Kingdom\\
$ ^{52}$Oliver Lodge Laboratory, University of Liverpool, Liverpool, United Kingdom\\
% ...
}
\end{flushleft}